\documentclass[aps]{revtex4-1}

\usepackage{graphicx}

\usepackage{graphics}

\newcommand{\be}{\begin{equation}}
\newcommand{\ee}{\end{equation}}
\newcommand{\nn}{\mbox{} \nonumber \\ \mbox{} }
\newcommand{\ba}{\begin{eqnarray}}
\newcommand{\ea}{\end{eqnarray}}

\newcommand{\E}{{\bf E}}
\newcommand{\B}{{\bf B}}

\renewcommand{\v}{{\bf v}}
\renewcommand{\k}{{\bf k}}
\newcommand{\n}{{\bf n}}
\newcommand{\p}{{\bf p}}

\newcommand\eg{\textit{e.g.,\ }}
\newcommand\cf{\textit{cf.,\ }}

\newcommand{\Bf}{{magnetic field}}

\newcommand{\Ef}{{electric  field}}

\newcommand{\EM}{electromagnetic}

\newcommand{\Sc}{Schwarzschild}

\newcommand{\Lf}{Lorentz factor}

\begin{document}

\title{Rotation of polarization  by  a moving gravitational lens}

\author{Maxim Lyutikov\\
Department of Physics and Astronomy, Purdue University, 
 525 Northwestern Avenue,
West Lafayette, IN
47907-2036, USA; lyutikov@purdue.edu\\
and \\
Department of Physics and McGill Space Institute, McGill University, 3600 University Street, Montreal, Quebec H3A 2T8, Canada}

\begin{abstract}
We present a simple prescription for the rotation of polarization produced by a relativistically  moving  gravitational lens, applicable to arbitrary deflection angle and arbitrary velocity of the lens.  When geometric optics is applicable, two independent components contribute to the total rotation of polarization:  (i) in the frame of the lens the polarization vector  experiences minimal rotation defined by the deflection angle (as measured by a set of remote observers, or no rotation if  defined in terms of parallel-propagated tetrad); (ii)  the effect of the motion of the lens on the polarization can be taken into account exactly using special relativistic  Lorentz transformation of polarization. The effects of the gravitational lensing are thus parametrized by the deflection angle of the null geodesics (not necessarily small) and the motion of the  lens  (not necessarily with velocities much smaller than that of light).
\end{abstract}
\maketitle

\section{Introduction}

In General Relativity (GR) the polarization vector is parallel-transported along the  light ray \citep{MTW}. 
%Thus, for  photon's trajectory in curved non-rotating space the polarization vector rotates minimally,  just to remain perpendicular to the null geodesics. 
 In {\Sc}ian (non-rotating) metric the final direction of polarization is uniquely determined by the initial polarization and the total deflection angle of the lens, \S \ref{stationary}. (In rotating metric there is additional contribution from the spin parameter \cite{1957SPhD....2..226S,1988PhRvD..38..472I}.) 

Moving gravitational lens  does lead to the rotation of polarization, \eg Ref. \cite{2002PhRvD..65f4025K}. In Ref. \citep{2017arXiv170108243P}  the GR principles behind this observer-dependent rotation of polarization were discussed. In this paper we point out that  the rotation of polarization due to the relative velocity of the lens and the observer is a special relativistic effect \citep{1972NPhS..240..161C,1979ApJ...232...34B,2003ApJ...597..998L} and thus  can be taken into account exactly, to any value of the velocity (and not only much smaller than the speed of light), \S \ref{app1}.

\section{Polarization rotation due to stationary {\Sc}ian lens}
\label{stationary}

For completeness, let us  first discuss the transformation of polarization by a non-rotating ({\Sc}ian) lens. The discussion below outlines  the concept of  parallel-transport
of polarization. 
The method we use to calculate polarization is somewhat different from what is usually done. For example, in Ref. \cite{2002PhRvD..65f4025K} the polarization transfer was done using  invariant relativistic formulation and then projecting  the 4-D results onto the 3-D space of an
observer. Instead, we define the observers from the beginning -  calculations  of the
polarization vector are done in an absolute 3D space. This method is simpler for the problem of moving lenses, when both the emitter and the observer are located far away from the source of gravity.

 Let  the initial direction of propagation and polarization in the frame of the lens be $\n_1'$ and $\p_1'$, $\n_1 '\cdot \p_1' =0$, and  the final states are
 $\n_2'$ and $\p_2'$, $\n_2 '\cdot \p_2' =0$ (primes denote quantities measured in the frame of the lens). The relation between $\n_1'$ and  $\n_2'$ is given by the GR bending of light \citep{MTW}. We would like to find final polarization $\p_2 '$ in terms of $\n_1 '$,  $\n_2'$ and  $\p_1'$.
 
 Introducing unit vector
\be
\k_r =  \frac{\n_1' \times \n_2' }{|\n_1' \times \n_2'|},
\ee
the polarization will be rotated from  $\p_1'$ around the axis $\k_r$ by the angle $\sin \theta_{rot} =  |\n_1' \times \n_2'|$. 
Rotation matrix around a direction $\k_r$ is 
\be
T_r = (1-\cos  \theta_{rot} ) \k_r \times \k_r + \cos  \theta_{rot}  {\bf I} + \sin _{rot} \theta {\k_r} _\times
\ee
where ${\k_r} _\times$ is the tensor product.

For example, if  the initial direction of propagation is $\n_1'=\{0,0,1\}$, polarization is  $\p_1'= \{1,0,0\}$ and the final direction is 
$\n_2 = \{ \cos \phi \sin \theta,  \sin \phi \sin \theta, \cos  \theta\}$ ($\theta$ is the deflection angle that can be calculated using GR for any photon trajectory), we find
\ba &&
\k_r =\{-\sin \phi ,\cos \phi ,0\}
\nn &&
\theta_{rot} = \theta
\nn &&
T_r = 
\left(
\begin{array}{ccc}
 (1-\cos \theta ) \sin ^2\phi +\cos \theta  & -(1-\cos \theta ) \sin \phi  \cos
   \phi  & \sin \theta  \cos \phi  \\
 -(1-\cos \theta ) \sin \phi  \cos \phi  & (1-\cos \theta ) \cos ^2\phi +\cos
   \theta  & \sin \theta  \sin \phi  \\
 -  \sin \theta  \cos \phi  & -\sin \theta  \sin \phi  & \cos \theta  \\
\end{array}
\right)
\nn &&
\p_2' =\left\{\cos \theta  \cos ^2\phi +\sin ^2\phi , - (1-\cos \theta ) \sin \phi  \cos
   \phi , - \sin \theta  \cos \phi \right\}
   \label{p2}
   \ea
   It may be verified that $\p_2 ' \cdot \n_2'=0$.
   (The statement that the polarization is not rotated during lensing by a stationary non-rotating lens, implies that the polarization is  just rotated by the above amount, and nothing  else.)
   
   For example, for  the photon's motion in the $x-z$ plane, $\phi =0$, the new direction of polarization is  $\p_2' = \{\cos \theta,0, - \sin \theta\}$, while for the photon's motion in the $y-z$ plane, $\phi = \pi/2$,  
     the polarization remains along $x$ direction, $\p_2' = \{1,0,0\}$. For small deviation angles $\theta \ll 1$, 
    $\p_2 '\approx  \{ 1,0, - \theta \cos \phi \}$. 

\section{Relativistic transformation of polarization}
\label{app1}

Let us next  derive explicit relations for the Lorentz transformation of polarization.
  Let a photon in some  frame be defined by a triad of unit vectors  along \Ef, \Bf\ and the direction of propagation ${\bf e}'$ , ${\bf  b}'$, $  {\bf  n}'$, so that ${\bf n}' = {\bf e}' \times {\bf  b}'$. The easiest way to derive relativistic transformation of polarization is using relations for the aberration of light,
\be
{\bf n} = \frac{{\bf n}'+ {\bf v} \gamma \left(1+ \frac{\gamma}{1+\gamma} (\bf{n}' \cdot \bf{v}) \right)}{\gamma (1+(\bf{n}' \cdot \bf{v})) }
\ee
  and transformation of the  \EM\ fields,
  \ba && 
  \B=\gamma \B' - \frac{\gamma^2}{\gamma+1} (\B' \cdot {\bf v}) {\bf v} - \gamma {\bf v} \times \E'
  \nn &&
   \E=\gamma \E' - \frac{\gamma^2}{\gamma+1} (\B' \cdot {\bf v}) {\bf v} + \gamma {\bf v} \times \B'
   \ea
Using $\E'=- {\bf n}' \times \B'$, for the unit vector along \Bf\ we find
\ba &&
{\bf b}=\frac{ {\bf q}'}{(1+\gamma) (1+({\bf n}' \cdot {\bf v})) }
\nn &&
{\bf q}'= {\bf b}' +{\bf v} \times ( {\bf b}' \times {\bf n}' )- \frac{\gamma}{\gamma+1} ({\bf b}' \cdot{\bf v}){\bf v}
%\nn && ={\bf b}' - \frac{({\bf b}' \cdot{\bf v}) }{(1+\gamma) (1+({\bf n}' \cdot {\bf v}) } \left( (1+\gamma) {\bf n}' + \gamma {\bf v} \right)
\label{polari}
\ea

Thus, the new direction of polarization (unit vector along ${\bf e}$) is 
\be
{\bf e}=
{  {\bf n} \times {\bf q}' \over 
\sqrt{ q^{\prime 2} - ( {\bf n} \cdot {\bf q}')^2} } \mbox{,}
\label{eeee}
\ee
(\cf \cite{2003ApJ...597..998L}). 
This solves in general the polarization transformation.  (In Appendix \ref{app2} we discuss how general relations for polarization transformation agree with relativistic  transformation of acceleration for dipolar-type emission.)

\begin{figure}[h!]
 \centering
 \includegraphics[width=.49\columnwidth]{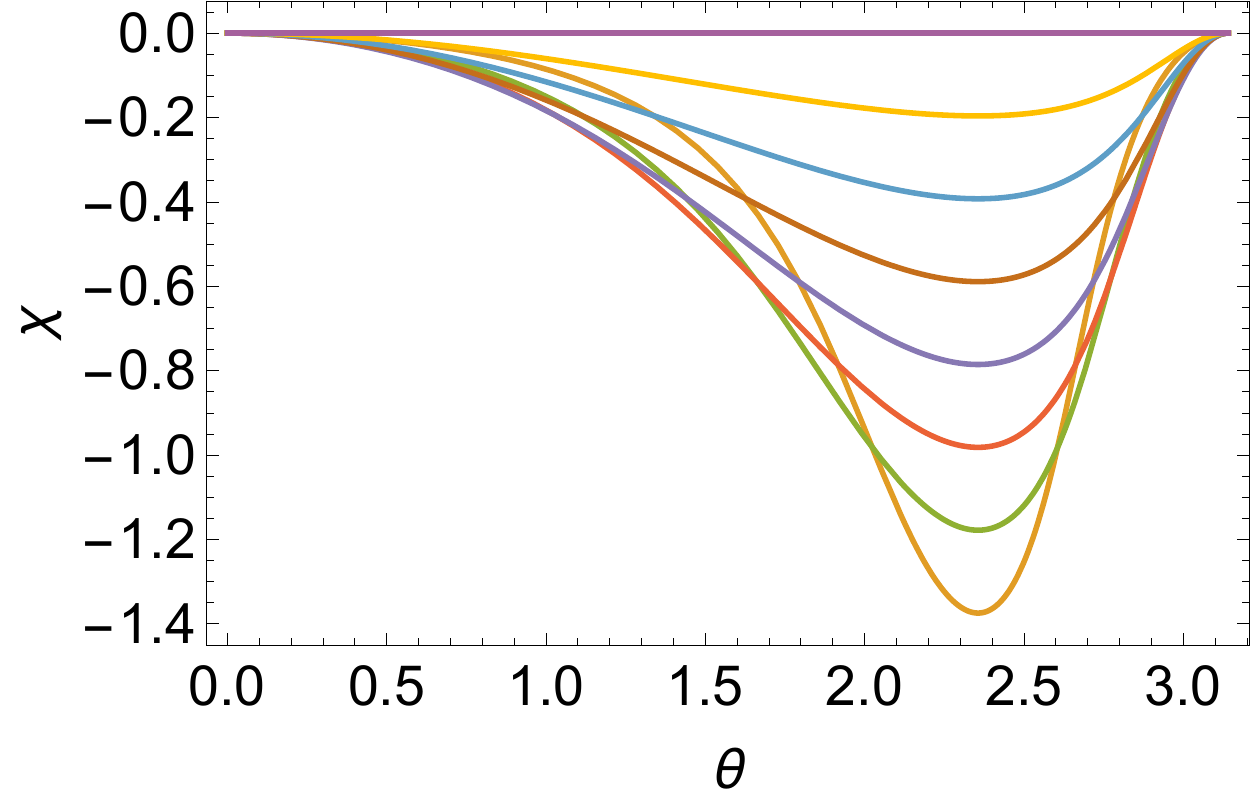}
 \includegraphics[width=.49\columnwidth]{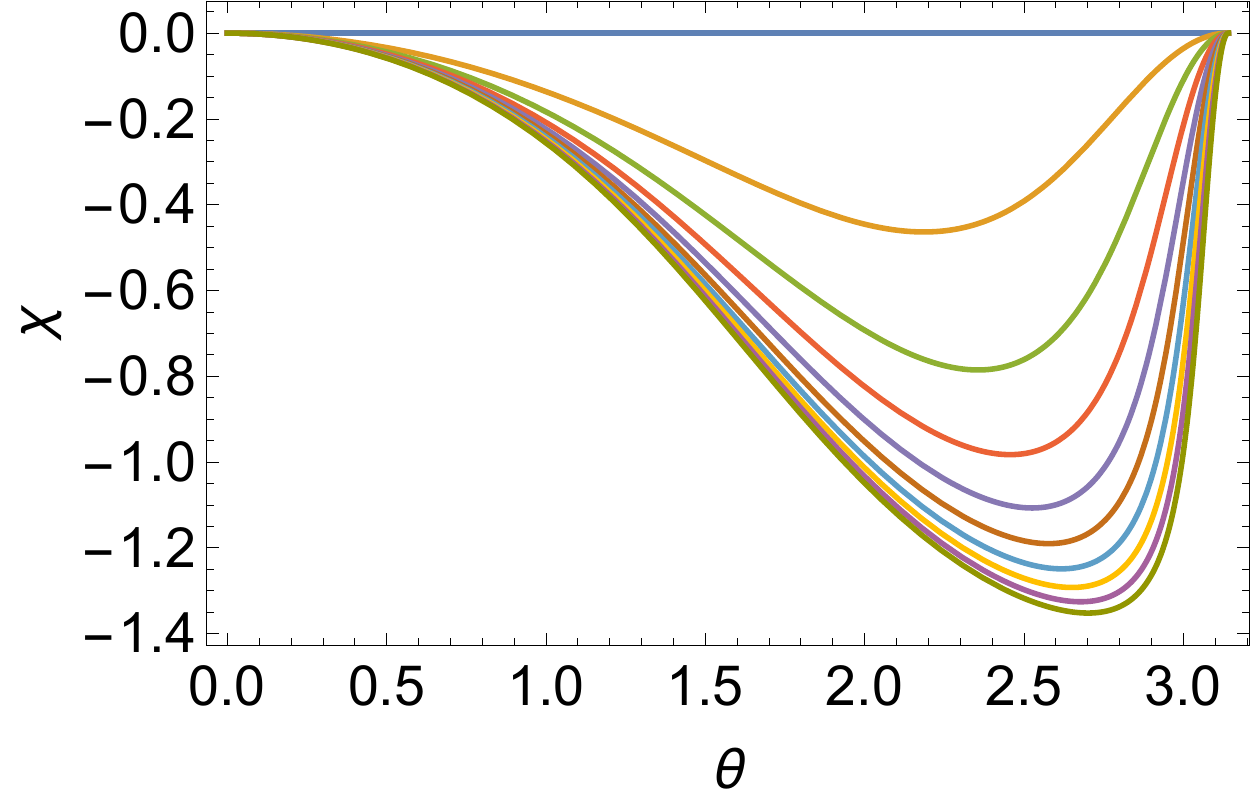}
  \caption{Rotation of the polarization angle $\chi$ as function of the direction of the velocity $\theta$ with respect to the photon propagation in the rest frame. 
  {\it Left Panel}: $\gamma=2$, different position angles $\phi=0, \, \pi/16\,  ... \, \pi/2$. {\it Right Panel}: $\phi=\pi/4$, different \Lf, $\gamma=1,\,2, \, ... \, 10$. (Only  a fraction of the curve is revealed to the observer through the generally narrow radio beam.)}
 \label{chioftheta}
\end{figure}

For example, if initially the photon propagates along $\hat{z}$ direction and is polarized along $\hat{x}$ direction, and the relative velocity is $ {\bf v}= \{\sin \theta_v  \cos \phi_v ,\sin \theta _v \sin \phi _v,\cos \theta _v\} \beta $,  the new direction of propagation and the polarization angle are, Fig. \ref{chioftheta}
\ba &&
{\bf n}_{\left(
\begin{array}{c}
x\\
y
\end {array}
\right)}
= \frac{\beta  \sin \theta _v (1+ \gamma(1+\beta    \cos \theta_v )}{(\gamma +1) (1+\beta  \cos \theta_v )} 
\left(
\begin{array}{c}
\cos \phi _v\\
\sin \phi _v
\end {array}
\right)
\nn &&
n_z=\frac{\beta  \gamma  \cos \theta _v +(\gamma -1) \cos ^2\theta_v +1}{\beta  \gamma  \cos \theta _v +\gamma }
   \nn &&
   \tan \chi = \frac{{\bf e}\cdot \hat{y}}{{\bf e}\cdot \hat{x}}= 
-\frac{\sin ^2\theta _v \sin \phi _v \cos \phi _v  (1+\beta  \cos \theta _v)}{\frac{\beta  (\gamma +1) \cos (\theta_v
   )+1}{\gamma -1}+\sin ^2\theta  _v \sin ^2 \phi  _v (1+\beta  \cos \theta _v)+\beta  \cos ^3\theta _v+\frac{(2 \gamma +1)
   \cos ^2\theta_v}{\gamma }}
\ea

In the limit $\gamma \rightarrow \infty$,
\be
\tan \chi =  -\frac{\sin ^2\theta  _v \sin \phi _v \cos \phi _v}{\sin ^2\theta_v  \sin ^2\phi_v +\cos ^2\theta_v +\cos \theta_v }
\ee
Thus, there is no rotation of polarization for special cases of motion along the photon direction, $\theta_v =0$, in the polarization plane,  $\phi_v=0$, or perpendicular to the polarization plane,  $\phi_v=\pi/2$.

\section{Polarization rotation due to moving lens}
Above in \S \ref{stationary} we derived polarization rotation due to stationary lens and in \S \ref{app1} we derived the Lorentz transformation of the polarization.
Combining  the two expressions  we can find polarization rotation due to (relativistically) moving lens. For practical purposes,  it is easies to start in the lens frame, and then (i) do a boost to lab frame to find the initial polarization in the lab frame; (ii) do GR-deflection and the corresponding rotation of polarization in the lens frame and do a boost to lab frame to find the final polarization in the lab frame.  

{\it Thus, the general procedure of polarization transfer by a moving lens involves Lorentz boost of  polarization vectors $\p_1'= \{1,0,0\}$ and $\p_2'$, Eq. (\ref{p2}), using transformation (\ref{eeee}).}  This procedure formally solves the problem of polarization rotation by a moving lens. (Alternatively, starting with wave vector and polarization in the laboratory frame, $\n_1, \, \p_1$, do a boost to the lens' frame and find directions and polarization, $\n_1', \, \p_1'$, do a radiative transfer in the frame of the lens to find  
$\n_2', \, \p_2'$, and then do a Lorentz boost back to the observer's frame to find $\n_2, \, \p_2$.)
Importantly, {\it  each step can be done exactly, without limitation of small angle deflection or small  relative velocity.}

As  practical applications are mostly limited to small lens velocities, next we find explicit relations for $\beta \ll 1$. 
If velocity of the lens in the lab frame is 
\be
\v= \{ \cos \phi_v \sin \theta_v,  \sin \phi _v \sin \theta _v, \cos  \theta_v\} \beta
\ee
then, keeping only linear terms in $\beta$, we find the initial polarization vector
\be
{\bf p}_0 = \{ 1,0, 0\} - \beta \{0,0, \cos \phi_v \sin \theta_v \}
\label{p0}
\ee
(\cf Ref. \cite{1989LNP...330...59B}, Eq. 1). There is no rotation if the lens is moving along the direction of propagation of the photon, $\theta_v=0$.

The final polarization vector is 
\ba &&
\p_2= \p_2'  + \beta \cos \mu \,  \hat{\bf r}
\nn &&
\hat{\bf r}= \{\sin\theta \cos\phi ,\sin\theta \sin\phi ,\cos \theta \}
\nn &&
 \cos \mu = 
 \sin\theta \cos\phi  \cos\theta _v-\sin\theta _v
   \left(\cos\theta \cos\phi  \cos \left(\phi -\phi _v\right)+\sin\phi  \sin
   \left(\phi -\phi _v\right)\right)
   \label{mu}
   \ea
   
   For example, consider a lens moving   in a direction perpendicular  to the direction of the photon propagation, $\theta_v = \pi/2$. We find then
   
   \begin{tabular}  {|c|c|c|}
   \hline
   & $ \p_0 $ & $ \p_2 $\\
   \hline
$ \phi =0, \, \phi_v =0  $ & $ \{1,0,-\beta \}$ & $\left\{\cos \theta  (\beta  \sin \theta +1),0,\beta  \cos ^2\theta -\sin (\theta
   )\right\} \approx  \{\beta  \theta +1,0,\beta -\theta \} $\\  \hline
$   \phi =\pi/2, \, \phi_v =0 $ & $\{1,0,-\beta \}$ & $ \{1,\beta  \sin \theta ,\beta  \cos \theta \} \approx \{1,\beta  \theta ,\beta \} $
   \\  \hline
$     \phi =\pi/2, \, \phi_v =\pi/2 $ & $ \{1,0,0 \}$ & $\{1,0,0 \} $\\  \hline
$       \phi =0, \, \phi_v =\pi/2 $ & $\{1,0,0 \}$ & $\{\cos \theta ,0,-\sin \theta \} \approx \{1,0,-\sin \theta \}$
\label{TT}
\end{tabular}

If the lens moves along the initial propagation of the photon, $\theta_v=0$, then $\p_0 = \{1,0,0 \} $ and 
\be
\p_2= 
\left\{
\begin{array} {cc}
\left\{\cos \theta -\beta  \sin ^2\theta ,0,\sin \theta  (-\beta  \cos (\theta
   )-1)\right\} \approx \{1,0, -(1+\beta) \theta \} \approx \{1,0,- \theta \}, & \mbox{for}\, \phi =\pi/2
\\
 \{1,0,0 \}, & \mbox{for}\, \phi =0
 \end{array}
 \right.
   \ee

   \section{Discussion}
   
   In this paper we solved exactly the polarization transfer by a moving gravitational lens by separating the motion of the lens and gravitation light bending in the frame of the lens. In the frame of the  spherical (non-rotating)  lens the final polarization direction is determined only by the total deflection angle in the frame of the lens (not necessarily small), while the motion of the lens can be taken into account exactly using Lorentz transformation of polarization. A typical value of the rotation angle is small $\sim \theta \beta$, where $\theta \approx  4 G M /(r c^2)$ is the deflection angle ($M$ is mass of the lens and $r$ is the impact parameter) and $\beta$ is the velocity of the lens \citep{2017arXiv170108243P}.

Our method of separating the GR effects in the frame of the  lens and special relativistic effects  due to the motion of the lens - both of which can be strong - can be applied to a more general case of rotating lens as well. Relations of \S \ref{stationary}  then need to be changed to include the effects of the lens' rotation, \eg Ref. \cite{2002PhRvD..65f4025K}.
   
   Possible applications of the effect include the Double Pulsar \citep{2004Sci...303.1153L}, where the line of sight passes $\sim 5 \times 10^8$ cm from the pulsar B \cite{2005ApJ...634.1223L};  so deflection angle is  $ \approx 1.5 \times 10^{-3}$.  The pulsar B moves with velocity $\sim 600$ km s$^{-1}$, $\beta =2 \times 10^{-3}$, so that the polarization rotation angle is 
   $\approx 3  \times 10^{-6}$. (Another case of nearly in-the-orbital-plane line of sight \cite{2010Natur.467.1081D} produces a much smaller effect.)
   
\begin{figure}[h!]
 \centering
 \includegraphics[width=.49\columnwidth]{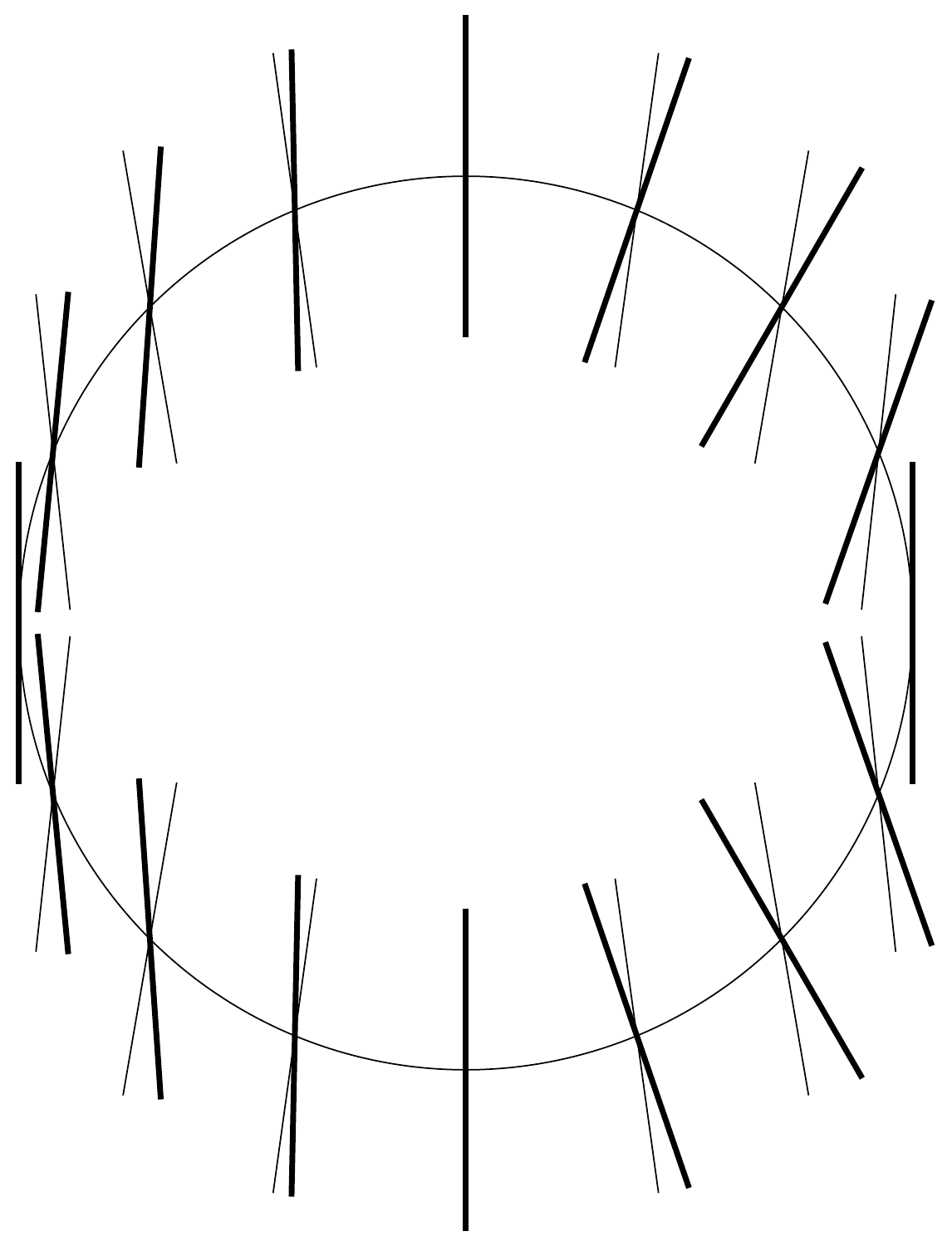}
  \caption{Polarization pattern produced by a moving strong lens as viewed by an observer. The source is  polarized along the vertical direction. Thin lines indicate polarization direction for stationary lens; thick lines indicate polarization direction for a lens moving horizontally,  in a  direction transverse to the initial photon propagation.  The circle represents the Einstein ring. The effects of motion are exaggerated for clarity.}
 \label{EinstringRingPolariz}
\end{figure}

   In case of cosmological gravitational lensing, the deflection angle for strong lensing is typically few arc-seconds, while the random velocity of the lens can be estimated as a  virial velocity in a large cluster of galaxies, $\beta \leq 10^{-2}$. Thus, the expected rotation of polarization is in the range of  milliarcseconds (see also Ref. \cite{2014PhRvL.112d1303D}).  The corresponding pattern for a lens propagating in a direction $\theta_v = \pi/2,\, \phi_v =\pi/2 $  in given in Fig. \ref{EinstringRingPolariz}.

% Finally, we note that the relative motion of the observer and the source of polarized emission does not lead to the E-B mixing  \cite{1997PhRvD..55.7368K}. It is the lensing that produces the B-mode - the motion of the lens keeps the total power in the B-mode. (Thus, for example, the motion of the Earth with respect to the CMB produces rotation of polarization, but does not lead to E-B mixing.)

   I would like to thank Jens Chluba,  Sergey Kopeikin,  Ue-Li Pen and  I-Sheng Yang for discussion. This work was supported by   NSF  grant AST-1306672 and DoE grant DE-SC0016369.

   \bibliographystyle{apsrev}
% \bibliographystyle{plain}
%\bibliographystyle{apj}
%\bibliography{~/Home/PulsarRadio/PulsarBib}
\bibliography{/Users/maxim/Home/Research/BibTex}
%\bibliography{/Users/maximlyutikov/Home/Research/BibTex}
%\bibliography{~/Home/Research/HallNS}
%\bibliography{/Users/maximlyutikov/Home/Research/HallNS/HallNS}

%\end{document}

\appendix 

\section{Dipolar emission: polarization transformation from transformation of  acceleration}
\label{app2}

In case of dipolar-type emission, the \Ef\ in the wave is perpendicular to the projection of the particle acceleration on the plane of the sky
\citep[][Eq. (67.6)]{LLII}:
\ba &&
{\bf e} \propto ( {\bf a} \times {\bf n}) \times {\bf n} \propto  {\bf a} - ({\bf a} \cdot {\bf n}) {\bf n}
\nn &&
{\bf b} \propto ( {\bf a} \times {\bf n})
\label{a1}
\ea
Lorentz transformation of acceleration reads \citep{2007AmJPh..75..615L}
\ba &&
{\bf a} = \frac{{\bf q}_a}{\gamma^2  (1-{\bf v}' \cdot {\bf v})^3 }
\nn &&
{\bf q}_a= {\bf a}' -\frac{\gamma}{1+\gamma} ({\bf a}' \cdot {\bf v}){\bf v} + {\bf v} \times ({\bf v}' \times {\bf a}')
\label{qa}
\ea
where ${\bf a}'$ is an acceleration of a particle moving with velocity ${\bf v}'$ in primed frame, ${\bf v}$ is the relative velocities of two frames,  Lorentz factor refers to the relative motion, $\gamma = 1/\sqrt{1-v^2}$.

The similarity between expressions for $ {\bf q}_a$ and ${\bf q}'$ , Eqns (\ref{qa}) and  (\ref{polari}), is highly revealing. 
For example, consider  emission by a highly relativistic particle moving along a curved trajectory. In this case ${\bf v}' \times {\bf a}'$ is directed along the normal to the orbital plane. Particle velocity in the primed frame is both along trajectory and along the direction of emission, ${\bf n}'  \parallel  {\bf v}' $. Thus,
\be
{\bf q}_a= {\bf a}' -\frac{\gamma}{1+\gamma} ({\bf a}' \cdot {\bf v}){\bf v} +{\bf v} \times  ({\bf n}' \times {\bf a}')
\ee
So that the direction of  acceleration (actually, the component perpendicular to the direction of wave propagation) transforms  exactly as the \Ef\ in the wave.

\end{document}